\documentstyle[psfig,referee]{mn}

\begin{document}

\title[On the relation between the coronal line emission]
{On the relation between the coronal line emission 
and the IR / X-ray emission
in Seyfert galaxies\thanks{Based on observations
with ISO, an ESA project with instruments funded by ESA
Member States (especially the PI countries: France, Germany, the Netherlands
and the United Kingdom) and with the participation of ISAS and NASA.}
}

\author[M.A. Prieto et al]{
        M. Almudena Prieto$^1,^2$,
        A. M. P\'erez Garc\'{\i}a$^2$\thanks{email:{\tt apg@ll.iac.es}}
        and J. M. Rodr\'{\i}guez Espinosa$^2$\\
%\offprints{A. M. P\'erez Garc\'{\i}a}
        $^1$European Southern Observatory, Garching, Germany\\
        $^2$Instituto de Astrof\'{\i}sica de Canarias, E-38200, 
La Laguna (Tenerife), Spain 
}

\date{Received..................; accepted...................}

\maketitle  
{\bf \small Accepted for publication in MNRAS}

\begin{abstract}

The relation between the X-ray, the coronal line and the infrared (IR)
 emissions in a sample of the brightest known Seyfert galaxies is analysed. A
close relationship between the absorption-corrected soft X-ray
emission and both the mid-IR and the coronal line emission is found
for the Seyfert type 2 objects in the sample. The coronal line and the
X-ray emissions are both main tracers of the central activity, hence their
relationship with the mid-IR emission points to nuclear energetic
 process  as the main responsibles of the heating of the
circumnuclear dust. On the other hand, the above relations do not seem to
 hold 
for the Seyfert type 1 discussed in the sample, at least when the
comparisons are done in a flux diagram.  This is partially because of the 
reduced number of objects of this type analysed in this work and 
the fact that the measured soft X-ray emission in Seyfert 1s is
systematically larger, by at least an order of magnitude, than that in
the Seyfert 2 counterparts.  Finally, the hard X-ray emission in the
studied sample appears unrelated to either the mid-IR or the coronal
line emission.
\end{abstract}

\section{Introduction}

Coronal lines arising in the spectra of active galactic nuclei (AGN)
 are unique tracers of the
pure central power mechanism. These lines require ionization potentials
(IP) beyond 50 eV and thus their study provide unique clues on  
the ultraviolet (UV) to soft X-ray  region of the AGN  ionizing spectrum, 
this being a region 
difficult to probe with observations because of the heavy absorption at
those energies. On the other hand, pure starbursts, where
[OIV] 25.9~$\mu$m  is  generally not present or very week (Genzel et
al. 1998; Lutz et al. 1998)
do show  photon energies below $\sim$ 50 eV. 

In a previous paper (Prieto, P\'erez Garc\'{\i}a \&
Rodr\'{\i}guez Espinosa  2001), the coronal lines fluxes
from [OIV]25.9~$\mu$m and [NeV]14.5~$\mu$m in the ISO (Infrared Space Observatory) spectra of a
sample of the brightest known Seyfert galaxies are found to be
directly related to the mid-IR continuum emission arising in these
objects. Furthermore, Perez-Garcia \&
Rodr\'{\i}guez Espinosa  (2001) report on the presence of 
two characteristic components in the IR continuum of Seyfert galaxies:
a warm component peaking at 16~$\mu$m and having a typical
temperature in the 120-170 K range,
and a cold component peaking at about 100~$\mu$m and having a characteristic 
temperature in
the 40-70 K range. In Prieto et al (2001), the coronal line emission is found 
 to be strongly 
correlated with the warm-IR component  but unrelated with the cold-IR one.
Because of the nuclear origin of coronal lines, the above
relations were interpreted as the  mid-IR emission being  due to
dust heated mostly by processes associated with  the  AGN. The 
far-IR however is a different component, most probably  due to dust heated by
star forming regions 
in the disc of these galaxies and by the interstellar radiation field, and
thus unrelated to the active nucleus.

In this paper, we explore the above  relations  further on the basis of
 available X-ray data for the galaxies in the sample. The reasoning
behind is as follows. The IP of $O^{2+}$ is about 50 eV and that of
$Ne^{3+}$ is roughly 100 eV, and thus a correlation between 
the [OIV] and [NeV] line
fluxes and the soft X-ray emission might be anticipated.  If, as argued
in Prieto et al (2001), the mid-IR emission 
is due to dust mostly heated within
the AGN nuclear region, a further relation between the mid-IR
continuum emission and the soft X-ray emission may also be expected.

The present sample contains the  brightest known 
Seyfert galaxies for which it was possible to obtain 
reliable measurements of 
the coronal lines [O IV]$\lambda
25.9\mu m$, [Ne V]$\lambda 14.3\mu m$ using ISO.
It includes
seven Seyfert type 2 and four Seyfert type 1 among which are the  
prototype sources N1068, N4151 and Circinus. The ISO coronal line
spectra of the sample are studied in Prieto \& Viegas (2000); the
mid-to-far IR continuum of the sample is studied in P\'erez Garc\'{\i}a \&
Rodr\'{\i}guez Espinosa (2001); the
relationship between the IR continuum and the coronal line emission is
discussed in Prieto et al. (2001).

\section{Observational data}

{\it Coronal line data}

The sample of Seyfert galaxies used in this work was originally
presented in Prieto \& Viegas (2000). All the galaxies were observed
with the ISOSWS (ISO-Short Wavelength Spectrometer) at the wavelengths
 of the coronal lines [O IV]$\lambda
25.9\mu m$, [Ne V]$\lambda 14.3\mu m$, [Mg VIII]$\lambda 3.02\mu m$ and
[Si IX]$\lambda 2.58\mu m$. Regardless of the Seyfert type, the
strongest coronal lines found in the ISO spectra of these objects are
the [OIV] and [NeV] lines (cf. Table 1 in Prieto \& Viegas). In the
present analysis, only the emission from these two lines is 
considered. The integrated fluxes correspond to the ISO-SWS
aperture of 20x33 arcsecs.

{\it Mid-IR continuum data}

Continuum fluxes at 16, 25, 60, 90, 120, 135, 180 and 200 $\mu m$ were
 measured with ISOPHT (ISO-Photopolarimeter) for all the 
galaxies in the sample by
P\'erez Garc\'{\i}a \& Rodr\'{\i}guez Espinosa (2001). 
Only the fluxes at 16, 25 and 60$\mu m$ are considered in this work. 
These data are
complemented with ground-based data at 10$\mu m$ existing for all the
sources. The latter are taken from compilations by Contini, 
Prieto \& Viegas (1998), Edelson (1978), Rieke \& Lebofsky (1978), 
Edelson, Malkan \& Rieke (1987)
and Maiolino et al. (1998).  The 10$\mu m$ fluxes correspond to integrated
fluxes within an aperture between 5 and 8.5 arcsec. The 16 and 25$\mu
m$ values correspond to integrated fluxes within a 120 or 180 arcsec
aperture, depending on the size of the objects. The 60 $\mu m$ data
were acquired with the C100 detector consisting of a 3x3 panoramic 
pixel array, each pixel projecting on 45 arcsec on the sky.

{\it Soft X-ray data}

For most of the galaxies, the soft X-ray emission,
namely  the 0.2-2.4 keV region, is primarily taken from the {\it ROSAT}
PSPC (Position Sensitive Proportional Counter). 
After searching in the literature, a large
diversity in the way  soft X-ray fluxes were derived and reported
was found.  As
soft X-ray fluxes are largely  dependent  on the absorption correction
applied,  to ensure homogeneity, whenever  PSPC
pointing observations were available,  X-ray fluxes  
were derived by ourselves.
In this case and due to the poor spectral resolution of the 
{\it ROSAT} PSPC, the general approach is to fit a single
power-law model corrected by absorption to the  PSPC spectra. We note that
among the various parameters derived from the fit, 
integrated fluxes are the least
dependent on the model adopted.
Free parameters in the fit were  the H column density N(H), the power-law spectral
index and the normalization factor.
 In all cases, statistically acceptable fits were obtained, with all
the fit parameters being derived within reasonable
constraints. 
 In those cases in which the H column density was found 
lower than the corresponding Galactic value, the latter was used as a
fixed parameter in the fit and the integrated X-ray flux derived
accordingly. The X-ray fluxes, their  1 sigma uncertainty and the N(H)
are given in Table 1.
Details of the fit for 
individual sources are given in the caption to Table 1.
 
In  the case of sources with known complex spectrum and/or morphology,
absorption-corrected fluxes were taken from the literature. References
for those and the adopted N(H) are given in the caption to Table 1.
In these cases, the error
associated with the X-ray flux represents the amplitude of the flux
variation between different measurements found in the literature.
Besides, some of the objects in the sample are known to be variable,
e.g. NGC~4151, NGC~5548. We expect to account for this effect by
including in the errors the maximum range of fluxes reported in the
literature.

{\it Hard X-ray data}
 
Absorption-corrected fluxes in the 2-10 keV, giving preference to {\it ASCA}
or {\it BESPOSAX} data, were taken from the literature for all the objects.
In general, when several values were found, the average was used and
the associated error reflects the amplitude of the variation.  Fluxes
and the sources for these data are also given in Table 1.  The sample
includes three known Compton thick sources: NGC~1068, Mrk~533 and
Circinus. In these cases, the reported hard X-ray fluxes may be
subjected to large uncertainties and should be taken with caution.
The large error bars associated to these sources accounts for the
large spread in values reported by different authors and that on the
model used.

\section{Results}

Fig. 1 compares the absorption-corrected soft X-ray emission with
the [OIV] and [NeV] coronal line fluxes respectively, as well as with
the 10, 16, 25 and 60$\mu m$ continuum emission for the galaxies in
the sample.  The comparison with the respective warm- and cold- IR
emission as defined in P\'erez Garc\'{\i}a \& Rodr\'{\i}guez Espinosa
(2001, see also section 1) is also shown in the last row.

Two facts are readily seen. First, there is a kind of separation in
behavior between the two Seyfert types; secondly, the soft X-ray fluxes
correlate rather well with the coronal line fluxes (Figure 1, first
row), particularly with [OIV], the line with better signal-to-noise in
the ISO spectra, and with the 10$\mu m$ continuum emission
fluxes (Fig.1, second row).  The correlation also holds when the soft
X-ray emission is compared with the progressively colder IR fluxes,
but degrades when ISO fluxes from 60 $\mu m$ on are considered
(Fig. 1, third row; to avoid redundancy the comparisons with fluxes
above 60 $\mu m$ are not shown). Similarly, as expected based on the
above, a positive trend appears when the comparison is done with the
warm IR flux whereas no correlation is seen with the cold IR fluxes
(Fig. 1, last row).

The intermediate Seyfert type 1.5 galaxy, NGC~4151, follows the trend
 defined by the Seyfert 2 type objects. However, the four Seyfert 1
 objects in the sample, NGC~5548, NGC~5033, Mrk~335 and Mrk~817,
 markedly depart from the Seyfert 2 behavior.  Indeed, the soft X-ray
 emission in these objects is larger by more than an order of
 magnitude than that measured in the Seyfert 2s.
The trends described above contrast with the pure scatter diagram that
arises when the comparison is done with the absorption-corrected hard
X-ray fluxes regardless of Seyfert type (Fig. 2). In this case, both
the coronal line emission and the mid-IR emission appear unrelated to
the hard X-ray emission.

For comparison purposes, Fig. 3 shows the same comparisons as in
Fig. 1 but in luminosity plots.  As expected, the trends seen in the
flux diagrams show are also seen in the luminosity diagrams. 
Note however that not correlation is present between the soft X-rays and the
60 $\mu m$ luminosities, as already revealed in the flux diagram. 
Also, the Seyfert 1
galaxies show larger X-ray luminosities than their Seyfert 2
counterparts for the same IR luminosity.  The few Seyfert 1s in the
sample appear rather scattered in the luminosity diagrams; there is a
hint for a positive trend in the comparison with the [OIV] coronal
line emission, yet the reduced number of S1s in the sample prevents
further conclusion.

To the best of our knowledge, only Penston et al. (1984) conducted a
similar study addressing the possible relationship between coronal
line and X-ray emissions in a large sample of AGN. These authors found
a pure scatter diagram when comparing in their case the optical [FeX]
6374~$\AA$  coronal line flux and the 2-10 keV flux. This result is
therefore fully in agreement with the results presented here.  No
comparisons involving soft X-ray and coronal line emissions are
reported in the literature and thus, the results presented here are
unique on that respect.

Many authors, however, have addressed the study of the soft X-ray to IR
relation in Seyfert galaxies on the basis of much larger samples than
the one used here.  Results from those works are briefly compared with
 our results in the following.

In general, there is broad agreement on the lack of correlation
between the X-ray and the far-IR {\it IRAS} luminosities in Seyfert
galaxies (e.g. David, Jones \& Forman  1992; Green, Anderson \& Ward 1992);
 indeed a
degradation of the X-ray/ IR correlation with increasing IR
wavelength is postulated by Giuricin, Mardirossian \& Mezzetti
 (1995) and Danese et
al. (1992). This conclusion is thus in agreement with the results
presented here.

Focusing on the mid-IR vs soft X-ray relationship, there is a larger
diversity in the reported results.  Green et al. (1992) report of no
correlation in their sample, regardless of the Seyfert type when
comparing the {\it IRAS} 12~$\mu$m and the Einstein 0.5-4.5 keV
luminosities whereas Giuricin et al.(1995) report on a weak positive
correlation with the 10~$\mu$m emission in Seyfert 2 objects only;
Rush et al (1996) find a tight correlations between the {\it ROSAT} 0.1-2.4
keV and the 12~$\mu$m  {\it IRAS} luminosity for both Seyfert types but with
Seyfert 1s having relatively more soft X-ray emission than Seyfert
2s for the same mid-IR luminosity by 1-2 orders of magnitude. This
shift in X-ray luminosities is also found by Green et al (1992).  We
find clear positive trends for the Seyfert 2 objects when comparing
their soft X-rays with either the coronal line or the 10--25~$\mu$m
emissions. A similar trend is not seen in the Seyfert 1s although
this may be due to the small number of objects of that class in the
sample. As expected, the S1s in the sample show X-ray luminosities
larger by 1-2 order of magnitude than those in S2s, which make them
depart from the Seyfert 2 trend.  Thus, our results are supported by
those found by Rush et al (1996). We believe that part of the reason
for the discrepancy with the results of
 Green et al and Giuricin et al. 
resides on the much harder X-ray band pass used by those authors, 0.5-
4.5 keV. This band might include an important contribution from the
central non-thermal AGN emission, which may be smearing out the
underlaying mid-IR to soft X-ray relation, readily seen when
considering the softer 0.1-2.4 keV region.

Moving to the hard X-rays, Giuricin et al. (1995) find the 2-10 keV
emission weakly related to the 10~$\mu$m emission in their sample.
Interestingly, the authors also point out that the apparent correlation
becomes weaker and weaker as they restrict themselves to the nearer
objects.  As indicated above, Penston et al (1984) find no correlation
between the coronal line [FeX] flux and the hard X-ray fluxes. We find
no apparent trend between the hard X-ray emission and either the
coronal line or the IR continuum emissions.  Although the sample is
small, we note that the observed scatter in the diagrams extends over
five orders of magnitude in the X-rays in a sample of relatively
nearby objects - the farthest object being less than 200 Mpc
distance. Part, but not all, of the scatter may be introduced by the
three Compton-thick sources in the sample. In any case, the lack of
correlation appears to point into the direction indicated by Guiricin
et al., namely, the progressive weakness of the relation for the
nearer objects in their sample.

\section{Discussion}

The [OIV] and [NeV] coronal lines discussed herein imply ionization
potential of 50 eV and 100 eV respectively for the corresponding ions
to be produced. Regardeless of weather 
photoionization and/or shocks, the strength
of these line should depend, at least partially, on the available
energy budget beyond 50 eV. Considering that the energy required to
produce those high ionization species is very close to the soft X-ray
energies analysed here, a trend between the coronal line emission and
the soft X-ray emission is to be expected. This is precisely what is
observed for the Seyfert 2 objects in this sample.

The soft X-ray emission in the Seyfert 2 objects is also related to
the mid-IR emission of these galaxies. The relationship, however, is
stronger when the comparison is made with the ground-based 10$\mu m$
fluxes, which is somehow expected if one considers that the apertures
used in that case are about or smaller than 5-8 arcsecs. A progressive
degradation of that correlation is observed with increasing IR
wavelength. On the other hand, Prieto et al. (2001) find the ISO
mid-IR emission closely related to the coronal line emission arising
in the nucleus of these galaxies.  The strongest relation is seen as
well with the 10$\mu m$ emission and degrades progressively with
increasing IR wavelength. Putting all together, the general trend of
the mid-IR emission when compared with either the X-ray or the coronal
emission points out the central AGN environment and the energetic
process associated with it as the dominant heating source of the
circumnuclear dust.  Some dispersion in the above relations is 
expected, howevwer, as a result of the
 contribution from other sources unrelated with the
active region.  For example, we may expect an additional heating
contribution by circumnuclear star--forming regions, particularly at
the IR wavelengths caused by the large ISO apertures.
  In the X-rays, star formation activity, X-ray binaries and gas halos
can also contribute to the soft X-ray emission.  Yet, considering the
unique link between coronal line emission and AGN activity, the good
correlations found indicate that those contributions are at least not
dominant.

The few Seyfert 1 galaxies in the sample present soft X-ray emissions
larger by at least an order of magnitude than those measured in the
Seyfert 2 objects.  This difference, already known from previous
studies based on much larger samples of Seyfert galaxies (e.g. Green
et al. 1992; Rush et al 1996), make them already depart from the
overall trend shown by the Seyfert 2s in both the coronal-line- and
 versus the soft X-ray and the mid-IR versus  soft X-ray relations.
  None of the above relations seem to
apply to the few S1s in this sample; yet, some hint of a positive trend
in the luminosity plot, in particular when comparing the [OIV] and
soft X-ray emissions, is however apparent. Clearly, any
confirmation of the above results requires the analysis of a much
large number of S1 objects.  We would, however, like to note the
following.
 Within the framework of the unified Seyfert schemes, the observed
larger X-ray flux in Seyfert 1s indicates our preferential line
of sight to the central source of these objects, that being otherwise
suppressed or much absorbed in Seyfert 2s as a result of the presence of a
central obscuring structure -the putative disc/torus. A lack of
correlation in S1s could indicate that the dominant and perhaps
anisotropic central X-ray emission in these objects prevents us for
seeing the underlying, somehow mandatory, relation between the coronal
line and the soft X-ray emission. In Seyfert 2, being the primary
central emission absorbed, the correlation of the coronal emission
with the central reprocessed soft X-ray emission can readily be
seen. It is worth mentioning that the only intermediate Seyfert type
(1.5) in the sample, NGC~4151, fits perfectly within the Seyfert 2
trend in all the diagrams.

Coronal lines are formed in an intermediate region between the broad
and the narrow line region. These lines are seen in both Seyfert type
with similar strengths. Their correlation with the mid-IR emission
regardless of the Seyfert type, indicates that in both cases the
emission arise most likely from the outer regions of the torus.  The
fact that both Seyfert type also span similar ranges of mid-IR
luminosities lead Rush et al (1996) to
 similar conclusion.  Because the coronal line emission in S2s further
correlates with the soft X-rays, both emissions should be co-spatial
and therefore soft X-rays may well be produced at the outer regions of
the torus.

One of the most popular scenarios attempting to explain the soft X-ray
emission in Seyfert 2 galaxies is the electron-scattering of nuclear
X-rays into our line of sight (Miller,
 Goodrich \& Matthews 1991). An alternative scenario is the case
 in which the soft X-rays are bremsstrahlung emission from hot gas
 heated by shocks induced by the central radiation pressure (Viegas \&
 Contini 1994, Morse et al. 1996; Contini \& Viegas 2001).  The
modeling of the spectral
 energy distribution of Seyfert galaxies by Contini et al. (1998 and
 references therein) shows that the re-emission by dust is closely
 related to bremsstrahlung from hot gas in the vicinity of the
 nucleus. The temperature of the gas and that of the grains are found
 closely related to each other. In these models, which account for the
 combined effect of photoionization and shocks, the soft X-rays are
 mainly emitted by hot gas in the immediate post shock region. Mutual
 heating between the dust and the gas leads to the corresponding
 emission by dust in the near- and mid-IR. The observed relationship
 between the mid-IR emission and the soft X-ray emission in the
 Seyfert 2 objects analysed here provide support to this scenario. The
 comparison of the models predictions with the present correlations
 will be discussed in detail in Viegas et al (in preparation).

Regarding the hard X-rays, there is an apparent lack of correlation
between the hard X-ray emission and either the coronal line (in
agreement with Penston et al. 1974) or the IR continuum (in partial
agreement with Giuricin et al. 1996 who only find a week correlation).
The result may be surprising, considering that with the exception of
the Compton thick sources the measured hard X-rays are mostly primary
AGN emission and therefore a link with other AGN-related emission
processes may be expected. On the other hand, we first note that the
energies required to produce the high ionization coronal lines are
well below 2 keV; and secondly, the heating of the dust with energies
above 2 keV would either destroy the dust or shift the peak of the
emission to higher IR frequencies. In the latter case a positive trend
between hard X-ray and near IR emissions may be expected and indeed
that seems to be observed, at least among Seyfert 1 galaxies
(e.g. Kotilanien et al 1992).  Clearly, the present analysis on a much
large sample of Seyfert galaxies could set more firm conclusions on
the results discussed.

\newpage
\begin{table}
\begin{tabular}{llllllll}
\hline
Object & Type& v& F(0.2-2.4keV)& errF(0.2-2.4keV)& N(H) &F(2-10 keV)& errF(2-10
keV)\\
 & & Km s$^{-1}$& erg cm$^{-2}$ s$^{-1}$& & $10^{21}cm^{-2}$& erg cm$^{-2}$ s$^{-1}$&\\
\hline 
\hline
 NGC5548 & 1     &5152 & 2.4$\times10^{-11}$     & 1.5$\times10^{-11}$& 0.16$\pm$0.01&6.7$\times10^{-10}$       &
 1.1$\times10^{-10}$\\
 NGC5929 & 2     &2490 & 2.49$\times10^{-13}$     & 1.3$\times10^{-14}$& 0.58$\pm$0.15& $\le$7.9$\times10^{-12}$ &\\
 Mrk817  & 1     &9436 & 4.04$\times10^{-12}$     & 2.3$\times10^{-13}$& 0.06$\pm$0.03&                  &\\
 Mrk335  & 1     &7688 & 7.00$\times10^{-11}$     & 4.0$\times10^{-12}$& 0.41 fix& 1.2$\times10^{-11}$      &
 4.0$\times10^{-11}$\\
 Mrk266  & 2     &8360 & 9.80$\times10^{-13}$     & 9.0$\times10^{-13}$& 0.6$\pm$0.27& $\le$8.0$\times10^{-13}$ &\\
 Mrk533  & 2     &8670 & 3.40$\times10^{-13}$     & & 1.8fix &  5.0$\times10^{-13}$         &\\
 Mrk334  & 2     &6582 & $\le$2.80$\times10^{-13}$&  &  0.44 fix & $\le$1.3$\times10^{-11}$&\\
 NGC1144 & 2     &8648 & 1.10$\times10^{-13}$    & 8.0$\times10^{-14}$ & 0.5 fix & $\le$1.2$\times10^{-11}$ &\\
 NGC5033 & 1     &876  & 8.0$\times10^{-12}$     & 6.0$\times10^{-12}$& 0.25$\pm$0.07 & 5.5$\times10^{-12}$      &\\
 CIRCINUS& 2     &436  & 1.60$\times10^{-11}$    & 2.0$\times10^{-12}$& 0.5$\pm$0.04 &7.4$\times10^{-11}$      &
 1.0$\times10^{-10}$\\
 NGC1068 & 2     &1140 & 5.60$\times10^{-11}$    & 2.4$\times10^{-11}$& 2.5$\pm$0.7 & 3.5$\times10^{-12}$      &
 8.0$\times10^{-12}$\\
 NCG4151 & 1.5   &980  & 4.00$\times10^{-12}$    & 1.0$\times10^{-12}$& $\sim49\pm2$ &2.7$\times10^{-10}$      &
 1.0$\times10^{-10}$\\
\hline 
\end{tabular}
\caption{{\footnotesize Mrk~335: Soft X-ray flux from Rush et al (1996) corrected by the
Galactic absorption.  The absorption-corrected hard X-ray flux from
George et al (1998).  NGC~5033: average soft X-ray flux from Polleta
et al (1996) corrected by the Galactic absorption. 
Absorption-corrected hard X-ray flux is from Bassani et al (1999).
Mrk~533: Soft X-ray flux from Polleta et al (1996) corrected by the
Galactic absorption. The {\it ROSAT}/PSPC pointing spectrum is very noisy:
we derive the same flux value from a single power-law fit to the PSPC
spectrum assuming spectral index $\alpha$ =-1 and Galactic
absorption.  Absorption-corrected hard X-ray flux from Bassani
et al. (1999).  NGC~1144: Soft X-ray flux derived in this work from a
power-law fit to the {\it ROSAT}/PSPC spectrum and Galactic absorption;
error represents one sigma uncertainty in the fit.  The
hard X-ray flux is an upper limit from Polleta et al (1996).
NGC~5929: Soft X-ray flux derived in this work from an absorbed
power-law fit to the {\it ROSAT}/PSPC spectrum; error represents one
sigma uncertainty. Hard X-ray is an upper limit from Polleta et al
(1996).  Mrk~266: Average soft X-ray flux from Wang et al. (1997)
derived from a range of N(H) values.  Only the flux from the South
component --identified as the AGN component by Davies et al (2000) --
is considered.   Hard X-ray flux from Polleta et al
(1996).  NGC~1068: absorption corrected soft X-ray flux from Guainazzi
et al (1998); the error accounts for Polleta et al (1996) value
corrected by the Galactic absorption. Absorption-corrected hard
X-ray flux from Bassani et al (1999); the error bar accounts for
the absorption-corrected flux from Guainazzi et al (1998).  NGC~4151:
absorption-corrected soft X-ray flux from Warwick et al (1996) and
Weaver et al (1994) (average value). Absorption-corrected hard
X-ray flux from Warwick et al (1996) and George et al (1998) 
(average value).  Circinus: absorption corrected soft X-ray
flux from Guainazzi et al (1998); error accounts for Polleta
et al (1996) value corrected by the Galactic absorption. 
Absorption-corrected hard X-ray flux from Guainazzi et al (1998);
error accounts for absorption-corrected flux from Matt et al
(1999).  Mrk~334: Soft X-ray flux derived in this work from {\it ROSAT}/PSPC
survey data assuming a power-law with photon index -2 and Galactic absorption.  The hard X-ray flux is an upper limit from
Polleta et al (1996).  Mrk~817: Soft X-ray flux derived in this work
from a single power-law fit to the {\it ROSAT}/PSPC spectrum corrected by
Galactic absorption;  errorrepresents one sigma
uncertainty.  No hard X-ray data available. NGC~5548:
Absorption-corrected soft X-ray flux from Iwasawa et al (1999) and
Nandra et al (1993) (average value). Absorption-corrected hard X-ray
 flux from George et al (1998). }
}

\end{table}
\newpage
\begin{figure*}
\psfig{figure=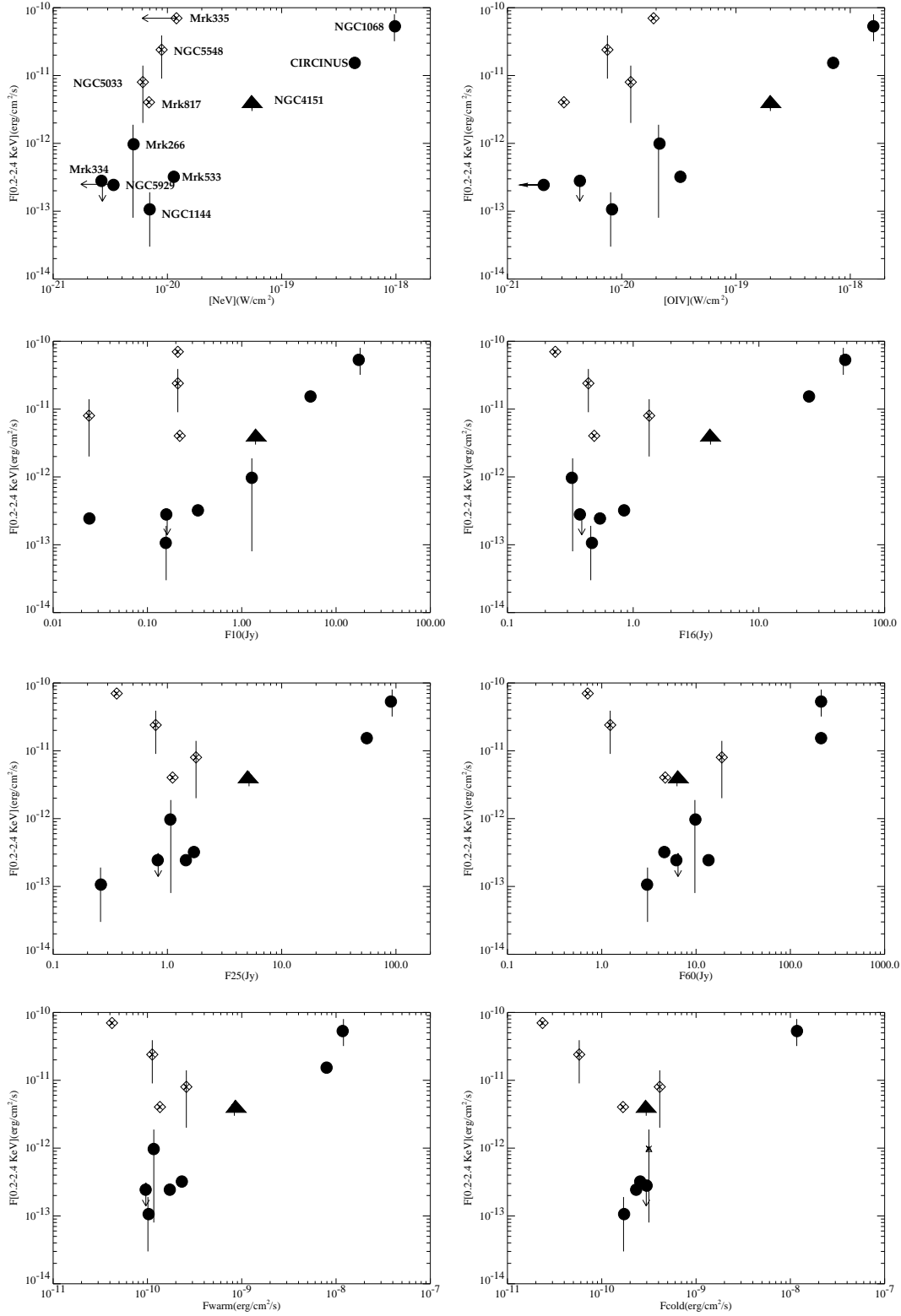,height=21cm}
\caption{The absorption-corrected soft X-ray emission for the galaxies in the
sample is compared with the [OIV] and [NeV] coronal line fluxes (first
row); with the 10, 16, 25 and 60$\mu m$ continuum emission (second and
third rows); and with the respective warm- and cold- IR emission as
defined in P\'erez Garc\'{\i}a \& Rodr\'{\i}guez Espinosa
 (2001) (last row).  Seyfert 1 galaxies
are represented by open diamonds, and the Seyfert 2 galaxies
are marked with filled circles. The intermediate type, NGC~4151 is
marked with a triangle.}  
\end{figure*}

\newpage
\begin{figure*}
\psfig{figure=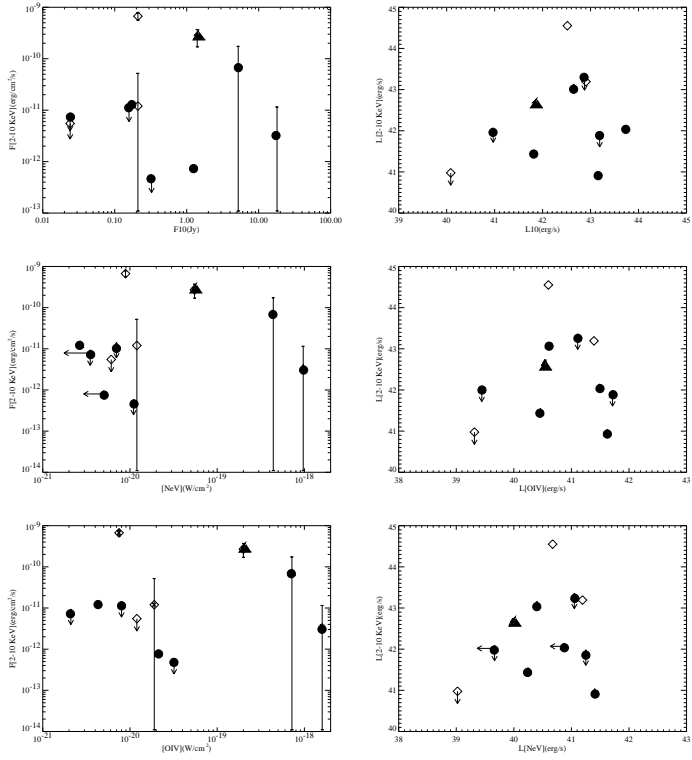,height=10truecm}
\caption{The absorption-corrected hard X-ray emission for the galaxies in the
sample is compared with the 10$\mu m$ emission and with [OIV] and [NeV]
coronal line emission. Fluxes are compared in the left column, luminosities 
in the right column. Symbols are
as in Fig. 1.  Notes on individual galaxies are given in the caption
to Table 1.}
\end{figure*}

\newpage
\begin{figure*}
\psfig{figure=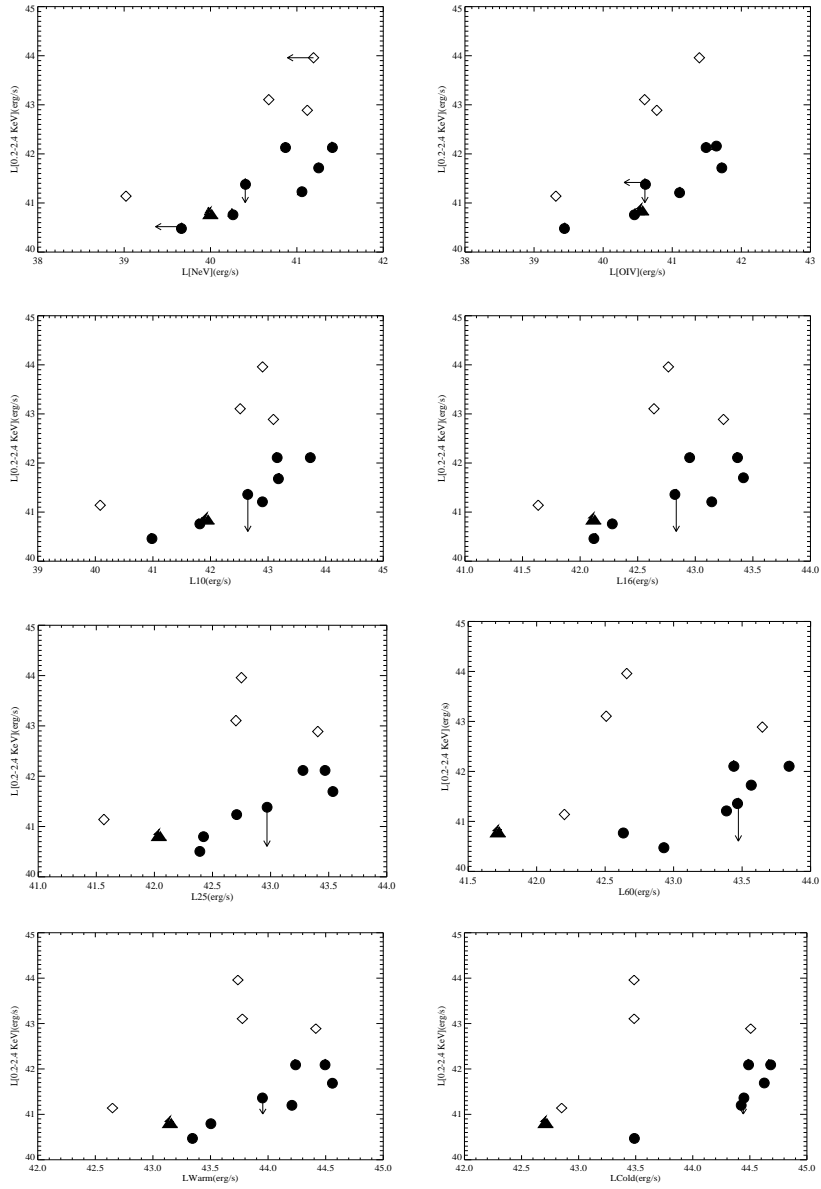,height=16cm}
\caption{As in Figure 1 but in luminosity plots,
Symbols as in Fig. 1. }
\end{figure*}

\end{document}